\documentclass[12pt]{iopart}

\begin{document}

\title[$PT$ symmetry and large-$N$ models]{$PT$ symmetry and large-$N$ models}

\author{Michael C. Ogilvie and Peter N. Meisinger}

\address{Department of Physics, Washington University, St. Louis, MO 63130
USA}
\ead{mco@physics.wustl.edu;pnm@physics.wustl.edu}

\begin{abstract}
Recently developed methods for PT-symmetric models can be applied to quantum-mechanical matrix and vector models. In matrix models, the calculation of all singlet wave functions can be reduced to the solution a one-dimensional PT-symmetric model. The large-N limit of a wide class of matrix models exists, and properties of the lowest-lying singlet state can be computed using WKB. For models with cubic and quartic interactions, the ground state energy appears to show rapid convergence to the large-N limit. For the special case of a quartic model, we find explicitly an isospectral Hermitian matrix model. The Hermitian form for a vector model with O(N) symmetry can also be found, and shows many unusual features. The effective potential obtained in the large-N limit of the Hermitian form is shown to be identical to the form obtained from the original PT-symmetric model using familiar constraint field methods. The analogous constraint field prescription in four dimensions suggests that PT-symmetric scalar field theories are asymptotically free.
\end{abstract}


\maketitle

\section{Introduction}

Since the initial discovery of $PT$ symmetry \cite{Bender:1998ke},
 there has been
considerable progress in expanding both the number of $PT$-symmetric
models and our knowledge of their properties \cite{Bender:2005tb,Bender:2007nj}.
However, models with continuous internal symmetry groups have not been extensively developed. Of course, the field theories relevant to modern particle physics have continuous symmetries, and it is natural to seek $PT$-symmetric models with
similar continuous symmetries. 
Here we review recent progress we have made in the
the construction and analysis of $PT$-symmetric models of scalars with $O(N)$ or $U(N)$ symmetry \cite{Meisinger:2007jx,Meisinger:2007qi}. 
Most of the results will deal with quantum mechanical models, usefully regarded as one-dimensional field theories. 
Of particular interest is the construction of the large-$N$ limit, as this has proven
to be a very powerful theoretical tool in the analysis of many different field theories. 

Hermitian matrix models appear in many contexts in modern theoretical physics,
with applications ranging from condensed matter physics to string
theory. Interest in the large-$N$ limit of matrix models was strongly
motivated by work on the large-$N_{c}$ limit of QCD  \cite{'tHooft:1973jz},
but interest today is much wider. For example, Hermitian
matrix quantum mechanics leads to a construction of two-dimensional quantum gravity
coupled to $c=1$ matter \cite{Kazakov:1988ch}.
It is surprising that the construction of $PT$-symmetric matrix
models is somewhat easier than the the construction of models with
a vector symmetry.
The matrix techniques pioneered in 
\cite{Brezin:1977sv} for Hermitian matrix quantum mechanics
can be extended to $PT$-symmetric matrix quantum mechanics.
In these models, the contours of functional integration
over the matrix eigenvalues are extended into the complex
plane.
The large-$N$ limit can then be taken in $PT$-symmetric matrix theories
just as in the Hermitian case. Quantities of interest such as the
scaled ground state energy and scaled moments can be calculated using
WKB methods. In the special case of a quartic potential with the {}``wrong''
sign, we use functional integration techniques to prove
 that the $PT$-symmetric
model is equivalent to a hermitian matrix model with an anomaly
 for all values of $N$, as
in the one-component case \cite{Bender:2006wt,Jones:2006et}. Interestingly,
the anomaly vanishes to leading order in the large-$N$ limit.

Although the construction
of $PT$-symmetric matrix models has proved to be relatively straightforward, 
the construction of $PT$-symmetric models with fields transforming
as vectors under $O(N)$ is more difficult technically.
Nevertheless, the development of scalar field theory models
with vector symmetry is crucial to the possible relevance of $PT$-symmetry
in particle physics. Only models with quartic interactions
have so far proved tractable. 
This progress on quartic models with $N$ components
is built upon recent work
on the relation of the
one-component $-\lambda x^{4}$
 model to its equivalent Hermitian form 
\cite{Bender:2006wt,Jones:2006et}, as well as
recent work on the relation of $O(N)$-symmetric Hermitian
models to one-component $PT$-symmetric models
\cite{Andrianov:2007vt}.
As in the single-component and matrix cases, 
the $PT$-symmetric model with $O(N)$ symmetry and quartic interaction
also proves to have a Hermitian form for all
values of $N$. 
The Hermitian form of the $PT$-symmetric $O(N)$ model
allows a technically straightforward
construction of the large-$N$ limit,
which can in turn be compared with simpler methods
that lead to essentially the same result at leading order.
The constraint-field method is particularly notable,
apart from its simplicity and familiarity,
because it generalizes to provide the form for the effective
potential of $PT$-symmetric scalar field theories
in the large-$N$ limit. This effective potential
in turn implies asymptotic freedom in
four dimensions, a property long
suspected to hold in a
renormalizable $PT$-symmetric
scalar field theory in four dimensions.

This paper is organized as follows: Section 2 develops the formalism required to
treat $PT$-symmetric matrix models, and section 3 analyzes the properties of the ground state of such models using WKB methods. Section 4 treats the special case
of the $TrM^{4}$ model, which has a simple Hermitian dual representation.
In section 5, a large class of $PT$-symmetric models with $N$ components are described, including vector
models with $O(N)$ symmetry. Section 6 shows that models in this class
have simple Hermitian dual
representations. In section 7, the large-$N$ limit of the $O(N)$ model
is derived using three different methods, while section 8 applies
one of these methods to $PT$-symmetric scalar field theories in the
large-$N$ limit.

\section{Formalism for Matrix Models}

The techniques for solving Hermitian matrix models are well-known.
The solution for all $N$ of the quantum mechanics problem associated with the
Euclidean Lagrangian\begin{equation}
L=\frac{1}{2}Tr\left(\frac{dM}{dt}\right)^{2}+\frac{g}{N}TrM^{4}\end{equation}
where $M$ is an $N\times N$ Hermitian matrix was first given by
Brezin {\it et al} \cite{Brezin:1977sv}.
The ground state $\psi$
is a symmetric function of the eigenvalues $\lambda_{j}$ of $M$.
The antisymmetric wave function $\phi$ defined by\begin{equation}
\phi\left(\lambda_{1},..,\lambda_{N}\right)=\left[\prod_{j<k}\left(\lambda_{j}-\lambda_{k}\right)\right]\psi\left(\lambda_{1},..,\lambda_{N}\right)\end{equation}
satisfies the Schrodinger equation\begin{equation}
\sum_{j}\left[-\frac{1}{2}\frac{\partial^{2}}{\partial\lambda_{j}^{2}}+\frac{g}{N}\lambda_{j}^{4}\right]\phi=N^{2}E^{(0)}\phi\end{equation}
where $E^{(0)}$is the ground state energy scaled for the large-$N$
limit. This equation separates into $N$ individual Schrodinger equations,
one for each eigenvalue, and the antisymmetry of $\phi$ determines
$N^{2}E^{(0)}$ as the sum of the $N$ lowest eigenvalues.

Here we solve the corresponding problem where the potential term is
$PT$-symmetric but not Hermitian. As shown by Bender and
Boettcher \cite{Bender:1998ke}, the
one-variable problem may be solved by extending the coordinate variable
into the complex plane. This implies that for $PT$-symmetric matrix
problems, we must analytically continue the eigenvalues of $M$ into
the complex plane, and in general $M$ will be normal rather than
Hermitian. We consider the Euclidean Lagrangian

\begin{equation}
L=\frac{1}{2}Tr\left(\frac{dM}{dt}\right)^{2}-\frac{g}{N^{p/2-1}}Tr\left(iM\right)^{p}\end{equation}
with $g>0$. Making the substitution $M\rightarrow U\Lambda U^{+}$,
with $U$ unitary and $\Lambda$ diagonal, we can write $L$ as
\begin{eqnarray}
L&=&\frac{1}{2}\sum_{j}\left(\frac{d\lambda_{j}}{dt}\right)^{2}+\sum_{j,k}\frac{1}{2}\left(\lambda_{j}-\lambda_{k}\right)^{2}\left(\frac{dH}{dt}\right)_{jk}\left(\frac{dH}{dt}\right)_{kj}\nonumber\\
&&-\frac{g}{N^{p/2-1}}\sum_{j}\left(i\lambda_{j}\right)^{p}
\end{eqnarray}
where\begin{equation}
\frac{dH}{dt}=-iU^{+}\frac{dU}{dt}.\end{equation}
In the analysis of conventional matrix models by Brezin {\it et al.}, a
variational argument shows that the ground state is a singlet, with
no dependence on $U$. Because the $\lambda_{j}$'s are
in general complex
for $PT$-symmetric theories, this
argument does not apply. However, in two cases we can prove
that the ground state is indeed a singlet: for $p=2$, which is trivial,
and for $p=4$, where the explicit equivalence with a hermitian matrix
model proven below can be used. Henceforth, we will assume that the
ground state is a singlet, but our results will apply in any case
to the lowest-energy singlet state.

We have now reduced the problem of finding the ground state to the
problem of solving for the first $N$ states of the single-variable Hamiltonian
\begin{equation}
H=\frac{1}{2}p^{2}-\frac{g}{N^{p/2-1}}\left(i\lambda\right)^{p}.\end{equation}
This Hamiltonian is $PT$-symmetric but in general not Hermitian.
The case $p=2$ is the simple harmonic oscillator. 
For $p>2$, the Schrodinger equation associated with each eigenvalue may
be continued into the complex plane as explained
in \cite{Bender:1998ke}.
We exclude the case $p<2$, where $PT$ symmetry is spontaneously broken
and the eigenvalues of $H$ are no longer real.

\section{Ground State Properties of Matrix Models}

As with Hermitian matrix models. the ground state energy is the sum
of the first $N$ eigenenergies of the Hamiltonian $H$. In the
large $N$ limit, this sum may be calculated using WKB.
A novelty of WKB for $PT$-symmetric models is the extension
of classical paths into the complex plane.
This topic has been treated extensively
for single-component models 
\cite{Bender:1998ke,Bender:1998gh}.

We define the Fermi energy $E_{F}$ as the energy of the $N$'th state
\begin{eqnarray}
N & = & \frac{1}{2\pi}\int dpd\lambda\,\theta\left[E_{F}-H(p,\lambda)\right]\end{eqnarray}
where the path of integration must be a closed, classical path in the
complex $p-\lambda$ plane. In order to construct the large-$N$ limit,
we perform the rescaling $p\rightarrow\sqrt{N}p$ and $\lambda\rightarrow\sqrt{N}\lambda$
yielding
\begin{equation}
H_{sc}(p,\lambda)=\frac{1}{2}p^{2}-g\left(i\lambda\right)^{p}\end{equation}
where the scaled Hamiltonian $H_{sc}$ is related to $H$ by  $H=NH_{sc}$.
We introduce a rescaled Fermi energy
 $\epsilon_{F}$ given by $E_{F}=N\epsilon_{F}$, which is implicitly
defined by 
\begin{equation}
1=\frac{1}{2\pi}\int dpd\lambda\theta\left[\epsilon_{F}-H_{sc}(p,\lambda)\right].
\label{eqn:one}\end{equation}
After carrying out the integration over $p$, we have\begin{equation}
1=\frac{1}{\pi}\int d\lambda\sqrt{2\epsilon_{F}+2g\left(i\lambda\right)^{p}}\theta\left[\epsilon_{F}+g\left(i\lambda\right)^{p}\right]\end{equation}
where the contour of integration is taken along a path between the
turning points which are the analytic continuation of the turning
points at $p=2$. This equation determines $\epsilon_{F}$ as a function
of $g$. 

We define a scaled ground state energy $E^{(0)}$ by
\begin{equation}
E^{(0)}_N=\frac{1}{N^{2}}\sum_{k=0}^{N-1}E_{k}.\end{equation}
The WKB result for the sum of the energies less than $E_{F}$ can be written
as \begin{equation}
\sum_{k=0}^{N-1}E_{k}=\frac{N^{2}}{2\pi}\int dpd\lambda\, H_{sc}(p,\lambda)\theta\left[\epsilon_{F}-H_{sc}(p,\lambda)\right]\end{equation}
so that in the large-$N$ limit $E^{(0)}_{\infty}$ is given by
\begin{equation}
E^{(0)}_{\infty}=
\frac{1}{2\pi}\int dpd\lambda\,H_{sc}(p,\lambda)\theta\left[\epsilon_{F}-H_{sc}(p,\lambda)\right]\end{equation}
The integration over $p$ is facilitated by using  equation (\ref{eqn:one}) to insert
a factor of $\epsilon_{F}$, giving
\begin{equation}
E^{(0)}_{\infty}=
\epsilon_{F}-\frac{1}{2\pi}\int dpd\lambda\left[\epsilon_{F}-H_{sc}(p,\lambda)\right]\theta\left[\epsilon_{F}-H_{sc}(p,\lambda)\right].\end{equation}
The
integral over $p$ then yields
\begin{equation}
E^{(0)}_{\infty}=\epsilon_{F}-\frac{1}{3\pi}\int d\lambda\left[2\epsilon_{F}+2g\left(i\lambda\right)^{p}\right]^{3/2}\theta\left[\epsilon_{F}+g\left(i\lambda\right)^{p}\right].
\label{eqn:E0}\end{equation}

The turning points in the complex $\lambda$ plane are\begin{equation}
\lambda_{-}=\left(\frac{\epsilon_{F}}{g}\right)^{1/p}e^{i\pi\left(3/2-1/p\right)}\end{equation}
\begin{equation}
\lambda_{+}=\left(\frac{\epsilon_{F}}{g}\right)^{1/p}e^{-i\pi\left(1/2-1/p\right)}\end{equation}
We integrate $\lambda$ along a two-segment, straight-line path
connecting the two turning
points via the origin \cite{Bender:1998ke}. 
Solving equation  (\ref{eqn:one})   for $\epsilon_{F}$, we
find\begin{equation}
\epsilon_{F}=
\left[\left(\frac{\pi}{2}\right)^{p}\left(\frac{\Gamma(3/2+1/p)}{\sin\left(\pi/p\right)\Gamma(1+1/p)}\right)^{2p}g^{2}\right]^{\frac{1}{p+2}},
\end{equation}
and solving (\ref{eqn:E0}) for the scaled ground state energy we have\begin{equation}
E^{(0)}_\infty=\frac{p+2}{3p+2}\epsilon_{F}
=\frac{p+2}{3p+2}\left[\left(\frac{\pi}{2}\right)^{p}\left(\frac{\Gamma(3/2+1/p)}{\sin\left(\pi/p\right)\Gamma(1+1/p)}\right)^{2p}g^{2}\right]^{\frac{1}{p+2}}.\end{equation}
 For $p=2$, this evaluates to $E^{(0)}=\sqrt{g/2}$ , in agreement
 with the explicit result for the harmonic oscillator.
 
It is very interesting to compare the large-$N$ result with results
for finite $N$. The low-lying eigenvalues for the Hamiltonian $p^{2}-(ix)^{p}$
have been calculated by Bender and Boettcher in \cite{Bender:1998ke} for the cases $p=3$ and
$p=4$; the case $p=2$ is trivial. We can use their results by noting
that the eigenvalues of our Hamiltonian $H$
 are related to theirs
by\begin{equation}
E_{j}=\frac{g^{2/(p+2)}}{2^{p/(p+2)}N^{(p-2)/(p+2)}}E_{j}^{BB}.\end{equation}
 Results for $p=3$ and $4$ and small values of $N$ are compared
with the large-$N$ limit in Table 1. The energies for finite values of $N$ rapidly approach the $N \to \infty$ limit. The approach to the limit appears monotonic in both cases, but with opposite sign.

\hfill
\begin{table}
\caption{The scaled ground state energy $E^{(0)}_N$ at $g=1$ for $p=3$ and $p=4$. The finite $N$ results obtained numerically rapidly approach the $N\rightarrow\infty$ limit obtained from WKB.}
\begin{center}
\begin{tabular}{|c||c|c|}
\hline 
N&
p=3&
p=4\tabularnewline
\hline
1&
0.762852&
0.930546\tabularnewline
\hline 
2&
0.756058&
0.935067\tabularnewline
\hline 
3&
0.75486&
0.935846\tabularnewline
\hline 
4&
0.754443&
0.936115\tabularnewline
\hline 
5&
0.754251&
0.936239\tabularnewline
\hline 
6&
0.754147&
0.936306\tabularnewline
\hline 
7&
0.754084&
0.936347\tabularnewline
\hline
8&
0.754043&
0.936372\tabularnewline
\hline
$\infty$&
0.753991&
0.936458\tabularnewline
\hline
\end{tabular}
\end{center}

\end{table}

The expected value of $\left\langle TrM\right\rangle $ for large $N$
is given by\begin{eqnarray}
\left\langle TrM\right\rangle =\sum_{j=0}^{N-1}\left\langle \lambda_{j}\right\rangle  & = & \frac{1}{2\pi}\int dpd\lambda\,\lambda\theta\left[E_{F}-H(p,\lambda)\right].\end{eqnarray}
Calculations of higher moments $\left\langle TrM^{n}\right\rangle $
are carried out in the same manner.
Upon rescaling, we find that $\left\langle TrM\right\rangle $ grows
as $N^{3/2},$ and the scaled expectation value is given by 

\begin{eqnarray}
\mu=\lim_{N\rightarrow\infty}\frac{1}{N^{3/2}}\left\langle TrM\right\rangle  & = & \frac{1}{2\pi}\int dpd\lambda\,\lambda\theta\left[\epsilon_{F}-H_{sc}(p,\lambda)\right]\end{eqnarray}
which reduces to \begin{equation}
\mu=\frac{1}{\pi}\int d\lambda\,\lambda\sqrt{2\epsilon_{F}+2g\left(i\lambda\right)^{p}}\theta\left[2\epsilon_{F}+2g\left(i\lambda\right)^{p}\right].\end{equation}
Using the same two-segment straight line path as before, we find that
%\begin{equation}
%\mu=\frac{\sqrt{2}}{\pi}\left(\epsilon_{F}\right)^{1/2+2/p}\left(\frac{1}{g}\right)^{2/p}\left(-2i\sin\left(\frac{2\pi}{p}\right)\right)\frac{\sqrt{\pi}\Gamma\left(1+2/p\right)}{4\Gamma\left(3/2+2/p\right)}\end{equation}
% which evaluates to
 \begin{equation}
\mu=-i\left(\frac{\pi}{2g}\right)^{\frac{1}{p+2}}
\frac{\cos\left(\frac{\pi}{p}\right)}{\sin\left(\frac{\pi}{p}\right)^{\frac{2}{p+2}}}\left[\frac{\Gamma(3/2+1/p)}{\Gamma(1+1/p)}\right]^{\frac{p+4}{p+2}}\frac{\Gamma\left(1+2/p\right)}{\Gamma\left(3/2+2/p\right)}.\end{equation}
For $p=2$, $\mu=0$, as expected for a harmonic oscillator. For $p>2$,
the expectation value $\mu$ is imaginary because $\left\langle \lambda_{j}\right\rangle $
for each eigenstate of the reduced problem is imaginary \cite{Bender:1998ke}.
For $p=3$, $\mu=-0.52006i$. For $p=4$, $\mu=-0.772539i$. In the
limit $p\rightarrow\infty$, $\mu$ goes to -i. This behavior is easy
to understand, because in this limit, the turning points become degenerate
at $-i$. 

\section{Special case of $TrM^{4}$}

For the case of a $TrM^{4}$ interaction, we can explicitly exhibit the equivalence
of the PT-symmetric matrix model with a conventional Hermitian quantum mechanical
system. As in the single-variable case, there is a parity-violating
anomaly, in the form of an extra term in the Hermitian form of the
Hamiltonian, proportional to $\hbar$. We show below that the anomaly
term does not contribute at leading order in the large-$N$ limit.

The derivation of the equivalence closely follows the path integral
derivation for the single-variable case \cite{Bender:2006wt,Jones:2006et}. 
The Euclidean Lagrangian is

\begin{equation}
L=\frac{1}{2}Tr\left(\frac{dM}{dt}\right)^{2}+\frac{1}{2}m^{2}Tr\, M^{2}-\frac{g}{N}TrM^{4}\end{equation}
 and the path integral expression for the partition function is\begin{equation}
Z=\int\left[dM\right]exp\left\{ -\int dt\, L\right\} .\end{equation}
Motivated by the case of a single variable, we make the substitution\begin{equation}
M=-2i\sqrt{1+iH}\end{equation}
where $H$ is an Hermitian matrix. Because $M$ and $H$ are simultaneously
diagonalizable, this transformation is tantamount to the relation\begin{equation}
\lambda_{j}=-2i\sqrt{1+ih_{j}}\end{equation}
between the eigenvalues of $M$ and the eigenvalues $h_{j}$ of $H$.
The change of variables induces a measure factor\begin{equation}
[dM]=\frac{[dH]}{Det[\sqrt{1+iH}]}\end{equation}
 where the functional determinant depends only on the eigenvalues
of $H$. The Lagrangian becomes\begin{equation}
L=\frac{1}{2}Tr\frac{(dH/dt)^{2}}{1+iH}-2m^{2}Tr\,(1+iH)-16\frac{g}{N}Tr\,(1+iH)^{2}\end{equation}
at the classical level. However, following \cite{Jones:2006et}, we note that in
the matrix case the change of variables introduces an extra term in
the potential of the form\begin{equation}
\Delta V=\sum_{j}\frac{1}{8}\left[\frac{d}{dh_{j}}\left(\frac{dh_{j}}{d\lambda_{j}}\right)\right]^{2}\end{equation}
which can be written as\begin{equation}
\Delta V=-\frac{1}{32}\sum_{j}\frac{1}{1+ih_{j}}=-\frac{1}{32}Tr\left(\frac{1}{1+iH}\right).\end{equation}
The partition function is now 
\begin{equation}
Z=\int\frac{\left[dH\right]}{\det\left[\sqrt{1+iH}\right]}
\exp\left[ -\int dt\, L\right] 
\end{equation}
where
\begin{eqnarray}
L&=&\frac{1}{2}Tr\frac{(dH/dt)^{2}}{1+iH}-2m^{2}Tr(1+iH)\nonumber\\
&&-\frac{16g}{N}Tr(1+iH)^{2}-\frac{1}{32}Tr\left(\frac{1}{1+iH}\right)
\end{eqnarray}
 We introduce a hermitian matrix-valued field $\Pi$ using the identity
 \begin{equation}
\frac{1}{\det\left[\sqrt{1+iH}\right]}=\int\left[d\Pi\right]
exp\left\{ -\int dt\, Tr\left[\frac{1}{2}\left(1+iH\right)\left(\Pi-\Pi_0\right)^{2}\right]\right\}
\end{equation}
where
$ \Pi_0 = \left(i\dot{H}+1/4\right)/\left(1+iH\right)$.
Dropping and adding appropriate total derivatives and integrating
by parts yields
\begin{equation}
Z=\int\left[dH\right]\left[d\Pi\right]
\exp\left[ -\int dt\, L'\right] 
\end{equation}
where
\begin{eqnarray}
L'&=&Tr\left[-2m^{2}(1+iH)-16\frac{g}{N}(1+iH)^{2}+\frac{1}{2}\left(1+iH\right)\Pi^{2}\right] \nonumber\\
&&+Tr\left[ \dot{\Pi}(1+iH)-\frac{1}{4}\Pi\right] 
\end{eqnarray}
The integration over $H$ is Gaussian, and
gives
\begin{equation}
Z=\int\left[d\Pi\right]exp\left\{ -\int dt\, Tr\left[\frac{N}{64g}\left(\dot{\Pi}^{2}-2m^{2}\Pi^{2}+\frac{1}{4}\Pi^{4}\right)-\frac{1}{4}\Pi\right]\right\} .\end{equation}
After the rescaling $\Pi\rightarrow\sqrt{32g/N}\Pi$ we have finally\begin{equation}
Z=\int\left[d\Pi\right]exp\left\{ -\int dt\, Tr\left[\frac{1}{2}\left(\dot{\Pi}^{2}-2m^{2}\Pi^{2}\right)+\frac{4g}{N}\Pi^{4}-\sqrt{\frac{2g}{N}}\Pi\right]\right\}.\end{equation}

This proves the equivalence of the $PT$-symmetric matrix model
defined by\begin{equation}
L=\frac{1}{2}Tr\left(\frac{dM}{dt}\right)^{2}+\frac{1}{2}m^{2}Tr\, M^{2}-\frac{g}{N}TrM^{4}\end{equation}
to the conventional quantum mechanics matrix model given by\begin{equation}
L'=\frac{1}{2}Tr\left(\frac{d\Pi}{dt}\right)^{2}-\sqrt{\frac{2g}{N}}Tr\Pi-m^{2}Tr\Pi^{2}+\frac{4g}{N}Tr\Pi^{4}.\end{equation}
 This equivalence implies that the energy eigenvalues of the corresponding
 Hamiltonians are the same. 
This could also be proven using the single-variable
equivalence for the special case of singlet states, but the functional
integral proof encompasses both singlet and non-singlet states at
once. The equivalence of these two models also allows for an easy
proof of the singlet nature of the ground state.
Standard variational arguments show that the ground state of
the Hermitian form is a singlet. The direct quantum mechanical equivalence
of the single-variable case is then sufficient to prove that the ground
state of the $PT$-symmetric form is also a singlet.

As in the single-variable case, there is a linear term
of order $\hbar$ appearing in the Lagrangian and Hamiltonian of the
Hermitian form of the model. This term represents
a quantum mechanical anomaly special
to the $TrM^{4}$ model. 
To determine the fate of the anomaly in the large-$N$ limit, 
we construct the scaled Hamiltonian 
of the Hermitian form in exactly the same way as for
the $PT$-symmetric
form. It is given by \begin{equation}
H_{sc}=\frac{1}{2}p^{2}-\frac{1}{N}\sqrt{2g}x-m^{2}x^{2}+4gx^{4},\end{equation}
indicating that the effect of the anomaly is absent in leading order
of the large-$N$ expansion. One easily checks for the $m=0$ case
that the Hermitian form without the linear term reproduces
the $PT$-symmetric prediction for $E^{(0)}_\infty$
at $p=4$.

\section{$O(N)$ Vector Models}

The analysis of the $O(N)$-invariant $PT$-symmetric model with a quartic
interaction is similar to that of the $TrM^{4}$ matrix model.
Consider a model with Euclidean Lagrangian given by\begin{equation}
L_{E}=\sum_{j=1}^{N}\left[\frac{1}{2}\left(\partial_{t}x_{j}\right)^{2}+\frac{1}{2}m^{2}x_{j}^{2}-\lambda x_{j}^{4}\right]-\frac{g}{N}\left(\sum_{j=1}^{N}x_{j}^{2}\right)^{2}\end{equation}
where $g$ and $\lambda$ are non-negative. When $g=0$, we have $N$
decoupled one-dimensional systems; for $\lambda=0$, we have a model
with $O(N)$ symmetry. When both $g$ and $\lambda$ are non-zero,
the model has only an $S_{N}$ permutation symmetry. From the standpoint
of $PT$ symmetry, the interaction terms can be considered as members
of a family of $PT$-invariant interactions\begin{equation}
-\lambda\sum_{j=1}^{N}\left(-ix_{j}\right)^{2p}-\frac{g}{N}\left(-\sum_{j=1}^{N}x_{j}^{2}\right)^{q}\end{equation}
which are invariant under $PT$ symmetry. This class of models
is well-defined for $p=q=1$, and must be defined for $p,q>1$ by
an appropriate analytic continuation of the $x_{j}$ as necessary
 \cite{Bender:1998ke}.

It is convenient to consider this model as a subset of a larger class
of models, with a Lagrangian of the form\begin{equation}
L_{E}=\sum_{j=1}^{N}\left[\frac{1}{2}\left(\partial_{t}x_{j}\right)^{2}+\frac{1}{2}m^{2}x_{j}^{2}\right]-\sum_{j,k=1}^{N}x_{j}^{2}\Lambda_{jk}x_{k}^{2}\end{equation}
The classical stability of the potential for large $x_{j}$ is governed
by the eigenvalues of $\Lambda$. For the model of particular interest
to us, \begin{equation}
\Lambda=\lambda I+gP\end{equation}
where $P$ is the one-dimensional projector\begin{equation}
P=\frac{1}{N}\left(\begin{array}{ccc}
1 & 1 & 1\\
1 & 1 & ..\\
1 & .. & ..\end{array}\right)\end{equation}
satisfying $P^{2}=P$. The decomposition $\Lambda=\lambda\left(I-P\right)+\left(g+\lambda\right)P$
shows that $\Lambda$ has one eigenvalue $g+\lambda$ and $N-1$ eigenvalues
with value $\lambda$. The eigenvalue $g+\lambda$ is associated with
variations in $\vec{x}^{2},$ i.e., variations in the radial direction.

\section{Equivalence of $PT$-symmetric vector models to Hermitian models}

We will analyze the case where all eigenvalues of $\Lambda$ are positive
using functional integration. With the substitution
\begin{equation}
x_{j}\rightarrow-2i\sqrt{c_{j}+i\psi_{j}}\end{equation}
familiar from the one-component case, $L_{E}$ becomes
\begin{eqnarray}
L_{E}&=&\sum_{j}\left[\frac{1}{2}\frac{\left(\partial_{t}\psi_{j}\right)^{2}}{(c_{j}+i\psi_{j})}-2m^{2}(c_{j}+i\psi_{j})\right]\nonumber\\
&&-16\sum_{jk}\Lambda_{jk}(c_{j}+i\psi_{j})(c_{k}+i\psi_{k}).
\end{eqnarray}
The generating function for the model is given by
 \begin{equation}
Z=\int\prod_{j}\frac{[d\psi_{j}]}{\sqrt{\det(c_{j}+i\psi_{j})}}\exp\left[-\int dt\left[L_{E}-\sum_{j}\frac{1}{32}\left(\frac{1}{c_{j}+i\psi_{j}}\right)\right]\right]\end{equation}
where the change of variables has generated both a functional determinant
and additional term, formally of order $\hbar^{2}$, in the action.
As pointed out in \cite{Jones:2006et}, both terms are required to obtain correct results
in the functional integral formalism.

The functional determinant may be written as 
\begin{eqnarray}
&&\prod_{j}\frac{1}{\det\left[\sqrt{c_{j}+i\psi_{j}}\right]}=\nonumber\\
&&\int\prod_{j}\left[dh_{j}\right]\exp\left\{ -\int dt\,\left[\frac{1}{2}\left(c_{j}+i\psi_{j}\right)\left(h_{j}-\frac{\dot{i\psi_{j}}+1/4}{c_{j}+i\psi_{j}}\right)^{2}\right]\right\} \end{eqnarray}
which introduces a new set of fields $h_{j}$. The derivation proceeds
as in the single-variable case. After integration by parts on the
$h_{j}\dot{\psi_{j}}$ terms, and adding and subtracting total derivatives,
the functional integral over the $\psi_{j}$ fields can be carried
out exactly. The integral is both local and quadratic, and requires
that the matrix $\Lambda$ have positive eigenvalues for convergence.
The result of this integration is \begin{equation}
Z=\int\prod_{n}\left[dh_{n}\right]\exp\left[-\int dt\, L_H \right] \end{equation}
where $L_H$ is given by
\begin{equation}
L_H =\frac{1}{64}\sum_{jk}\Lambda_{jk}^{-1}\left[\frac{1}{2}h_{j}^{2}+\dot{h}_{j}-2m^{2}\right]\left[\frac{1}{2}h_{k}^{2}+\dot{h}_{k}-2m^{2}\right]-\sum_{j}\frac{1}{4}h_{j}.
\end{equation}
After discarding total derivatives, we obtain
\begin{equation}
L_H =\frac{1}{64}\sum_{jk}\Lambda_{jk}^{-1}\left[\dot{h}_{j}\dot{h}_{k}+\frac{1}{4}\left(h_{j}^{2}-4m^{2}\right)\left(h_{k}^{2}-4m^{2}\right)\right]
-\sum_{j}\frac{1}{4}h_{j}
\end{equation}
 which gives the Hermitian form for our general $PT$-symmetric model
with $N$ fields.

In the particular case we are interested in, we have \begin{equation}
\Lambda^{-1}=\frac{1}{\lambda}\left(I-P\right)+\frac{1}{g+\lambda}P.\end{equation}
The Lagrangian may be written as
\begin{eqnarray}
L_{E}&=&\frac{1}{64\lambda}\sum_{j}\left[\dot{h}_{j}^{2}+\frac{1}{4}\left(h_{j}^{2}-4m^{2}\right)^{2}\right]-\frac{1}{4}\sum_{j}h_{j}\nonumber\\
&&-\frac{g}{64N\lambda\left(g+\lambda\right)}\left[\left(\sum_{j}\dot{h}_{j}\right)^{2}+\frac{1}{4}\left(\sum_{j}\left(h_{j}^{2}-4m^{2}\right)\right)^{2}\right].\end{eqnarray}
 It is helpful to immediately rescale all the fields as
 $h_{j}\rightarrow\sqrt{32\lambda}h_{j}$:
 \begin{eqnarray}
L_{E}&=&\sum_{j}\left[\frac{1}{2}\dot{h}_{j}^{2}+4\lambda\left(h_{j}^{2}-\frac{m^{2}}{8\lambda}\right)^{2}\right]-\sqrt{2\lambda}\sum_{j}h_{j}\nonumber\\
&&-\frac{g}{N\left(g+\lambda\right)}\left[\frac{1}{2}\left(\sum_{j}\dot{h}_{j}\right)^{2}+4\lambda\left(\sum_{j}\left(h_{j}^{2}-\frac{m^{2}}{8\lambda}\right)\right)^{2}\right].\end{eqnarray}
 At this point, the $S_{N}$ permutation symmetry is still manifest,
and it clear that the field $\sum_{j}h_{j}$ plays a special role.

In order to understand the strategy for rewriting the model in a form
in which the limit $\lambda\rightarrow0$ can easily be taken, it
is useful to work out explicitly the case of $N=2$ first. It is apparent
that a rotation of the fields will be desirable. We define suggestively
new fields $\sigma$ and $\pi$ given by\begin{equation}
\begin{array}{c}
h_{1}=\frac{1}{\sqrt{2}}\left(\sigma+\pi\right)\\
h_{2}=\frac{1}{\sqrt{2}}\left(\sigma-\pi\right)\end{array}.\end{equation}
 After some algebra and the rescaling\begin{equation}
\sigma\rightarrow\sqrt{\frac{g+\lambda}{\lambda}}\sigma\end{equation}
we arrive at
\begin{eqnarray}
L_{E}&=&\frac{1}{2}\dot{\sigma}^{2}+\frac{1}{2}\dot{\pi}^{2}-m^{2}\sigma^{2}-\frac{\lambda m^{2}}{g+\lambda}\pi^{2}+2\left(g+\lambda\right)\sigma^{4}+\frac{2\lambda^{2}}{g+\lambda}\pi^{4}\nonumber\\
&&+\left(8g+12\lambda\right)\sigma^{2}\pi^{2}-2\sqrt{g+\lambda}\sigma.
\end{eqnarray}
Notice the natural hierarchy between the masses for $\lambda\ll g$.
The $O(2)$ symmetric limit of the original $PT$-symmetric model
is obtained in the limit $\lambda\rightarrow0$, where we have\begin{equation}
L_{E}=\frac{1}{2}\dot{\sigma}^{2}+\frac{1}{2}\dot{\pi}^{2}-m^{2}\sigma^{2}+2g\sigma^{4}+8g\sigma^{2}\pi^{2}-2\sqrt{g}\sigma.\end{equation}
The field $\pi$ has no mass term, indicating its relation to the
angular degrees of freedom in the original Lagrangian. However, radiative
corrections generate a mass for the $\pi$ field via the the $\sigma^{2}\pi^{2}$
interaction. As in the one-component case, there is a linear anomaly term, but
only for $\sigma$.

We now turn to the more difficult case of the $\lambda\rightarrow0$
limit for arbitrary $N$. As before, we introduce a field $\sigma$
defined by\begin{equation}
\sigma=\frac{1}{\sqrt{N}}\sum_{j}h_{j}\end{equation}
as well as a set of $N-1$ fields $\pi_{k}$ with $k=1,..,N-1$ related
to the $h_{j}$ fields by a rotation so that $\sigma^{2}+\vec{\pi}^{2}=\vec{h}^{2}$.
Each field $h_{j}$ can be written as\begin{equation}
h_{j}=\frac{1}{\sqrt{N}}\sigma+\tilde{h}_{j}\end{equation}
 where $\sum_{j}\tilde{h}_{j}=0$. This property is crucial in eliminating
a term in $L_{E}$ which diverges as $\lambda^{-1/2}$ as $\lambda\rightarrow0$. 
The Lagrangian now can be written as
\begin{eqnarray}
L_{E}&=&\frac{1}{2}\dot{\sigma}^{2}+\sum_{j}\frac{1}{2}\dot{\pi}_{j}^{2}-m^{2}\left(\sigma^{2}+\vec{\pi}^{2}\right)+4\lambda\sum_{j}h_{j}^{4}\nonumber\\
&&+\frac{4}{N}\left(\frac{\lambda^{2}}{g+\lambda}-\lambda\right)\left(\sigma^{2}+\vec{\pi}^{2}-\frac{Nm^{2}}{8\lambda}\right)^{2}-\sqrt{2\lambda N}\sigma.
\end{eqnarray}
The rescaling $\sigma\rightarrow\sqrt{\left(g+\lambda\right)/\lambda}\sigma$
plus some careful algebra yields the $\lambda\rightarrow0$ limit as\begin{equation}
L_{E}=\frac{1}{2}\dot{\sigma}^{2}+\frac{1}{2}\dot{\vec{\pi}}^{2}-m^{2}\sigma^{2}+\frac{4g}{N}\sigma^{4}+\frac{16g}{N}\sigma^{2}\vec{\pi}^{2}-\sqrt{2gN}\sigma\end{equation}
 which agrees with our previous result for $N=2$ , and agrees with
the known result for a single degree of freedom if we take $N=1$
and drop the $\vec{\pi}$ field altogether. This is a Hermitian form
of the $PT$-symmetric anharmonic oscillator with $O(N)$ symmetry,
derived as the limit of a $PT$-symmetric model with $S_{N}$ symmetry.
The Hermitian form has several novel features. Note that both the
$S_{N}$ and $O(N)$ symmetries are no longer manifest, but there
is an explicit $O(N-1)$ symmetry associated with rotations of the
$\vec{\pi}$ field. As in the $N=2$ case, there is no mass term for
the $\vec{\pi}$ field. Furthermore, there is no $\left(\vec{\pi}^{2}\right)^{2}$
term, although there is a $\vec{\pi}^{2}\sigma^{2}$ interaction.
The anomaly term again involves only $\sigma$, and breaks the symmetry
$\sigma\rightarrow-\sigma$ possessed by the rest of the Lagrangian.
Analyzing the Lagrangian at the classical level, we see that if $m^{2}>0$
, the $\sigma$ field is moving in a double-well potential, perturbed
by the anomaly so that $\left\langle \sigma\right\rangle >0$.$ $
On the other hand, if $m^{2}<0$, $\sigma$ moves in a single-well
anharmonic oscillator, again with the linear anomaly term making $\left\langle \sigma\right\rangle >0$.
In either case, the $\vec{\pi}^{2}\sigma^{2}$ interaction will generate
a mass for the $\vec{\pi}$ field. All of this is consistent with
the association of $\sigma$ and $\vec{\pi}$ with the radial and
angular degrees of freedom, respectively, in the original $PT$-symmetric
model. % Revised
This equivalence between $PT$-symmetric and Hermitian forms may be compared to the results of \cite{Andrianov:2007vt},
where a somewhat different equivalence is derived. In that work, the generating function for a Hermitian $x^4$ theory is shown to be equivalent to a sum over generating functions for a class of single-component $PT$-symmetric models, with each element of the class representing a different angular momentum. As we discuss below in the context of the large-$N$ limit, both approaches lead to an anomaly term with a linear dependence on the angular momentum quantum number $l$.

\section{Large-$N$ Limit of Vector Models}

We will defer a more detailed discussion of this model for finite
$N$, and turn to its large-$N$ limit. One more rescaling $\sigma\rightarrow\sqrt{N}\sigma$
gives the Lagrangian\begin{equation}
L_{E}=\frac{N}{2}\dot{\sigma}^{2}+\frac{1}{2}\dot{\vec{\pi}}^{2}-Nm^{2}\sigma^{2}+4gN\sigma^{4}+16g\sigma^{2}\vec{\pi}^{2}-N\sqrt{2g}\sigma.\end{equation}
 We see that the anomaly term survives in the large-$N$ limit, unlike
the matrix model case \cite{Meisinger:2007jx}. After integrating over the $N-1$ 
components of the $\vec{\pi}$
field, we have the large-$N$ effective potential $V_{eff}$ for
$\sigma$:\begin{equation}
V_{eff}/N=-m^{2}\sigma^{2}+4g\sigma^{4}+\frac{1}{2}\sqrt{32g\sigma^{2}}-\sqrt{2g}\sigma.\end{equation}
It is striking that the anomaly term has virtually the same form as
the zero-point energy of the $\vec{\pi}$ field. The anomaly term
breaks the discrete $\sigma\rightarrow-\sigma$ symmetry of the other
terms of the Lagrangian, and always favors $\sigma\geq0$. The effective
potential has a global minimum with $\sigma$ positive for $m^{2}>3\,2^{1/3}g^{2/3}$.
For $m^{2}<3\,2^{1/3}g^{2/3}$, there does not appear to be a stable
solution with $\sigma>0$, and $\sigma=0$ is the stable solution
to leading order in the $1/N$ expansion. This change in the behavior
of the effective potential as $m^{2}$ is varied is not seen in the
corresponding Hermitian model \cite{Coleman:1974jh}, and indicates
a need for care in analyzing the model. Based on our preliminary analysis
of the Hermitian form for finite $N$, we believe that this behavior
is associated with the large-$N$ limit, and does not indicate a fundamental
restriction on $m^{2}$.

The large-$N$ effective potential was derived from a Lagrangian with
unusual properties, associated with the Hermitian form of the original
model. It is therefore surprising that, once the form of the large-$N$
effective potential is known, it can be derived heuristically in a
more conventional way. We start from the $O(N)$-symmetric Lagrangian\begin{equation}
L_{E}=\sum_{j=1}^{N}\left[\frac{1}{2}\left(\partial_{t}x_{j}\right)^{2}+\frac{1}{2}m^{2}x_{j}^{2}\right]-\frac{g}{N}\left(\sum_{j=1}^{N}x_{j}^{2}\right)^{2}\end{equation}
 and add a quadratic term in a constraint field $\rho$\begin{equation}
L_{E}\rightarrow L_{E}+\frac{g}{N}\left(\frac{2N\rho}{g}+\sum_{j=1}^{N}x_{j}^{2}-\frac{Nm^{2}}{4g}\right)^{2}\end{equation}
yielding\begin{equation}
L_{E}=\sum_{j=1}^{N}\left[\frac{1}{2}\left(\partial_{t}x_{j}\right)^{2}+4\rho x_{j}^{2}\right]+\frac{4N\rho^{2}}{g}-\frac{Nm^{2}\rho}{g}+\frac{Nm^{4}}{16g}.\end{equation}
If we integrate over $x_{j}$ in a completely conventional way, we
obtain the large-$N$ effective potential\begin{equation}
V_{eff}/N=\frac{4\rho^{2}}{g}-\frac{m^{2}\rho}{g}+\sqrt{2\rho}+\frac{m^{4}}{16g}.\end{equation}
This is essentially identical to our previous expression after identifying
$\rho=g\sigma^{2}$. However, we lack a fundmental justification for
this approach. We know that great care must be taken in specifying
the contour of integration in typical $PT$-symmetric models, yet
the $x_{j}$ fields were integrated over quite conventionally. If
this approach has validity, it seems likely that the choice of integration
contours for $\rho$ and $\vec{x}$ is crucial. However, only the
saddle point matters to leading order in $1/N$, so it is possible
for this heuristic derivation to be correct even though we lack a
direct, complete treatment of the original $PT$-symmetric model.

There is another approach to the effective potential that
sheds some
light on the role of angular momentum in $PT$-symmetric vector models.
Let us take as our starting point the Hamiltonian for the
$PT$-symmetric vector model after the introduction of
the constraint field:
\begin{equation}
H=\sum_{j=1}^{N}\left[\frac{1}{2}p_{j}^{2}+4\rho x_{j}^{2}\right]+\frac{4N\rho^{2}}{g}-\frac{Nm^{2}\rho}{g}
\end{equation}
where for simplicity we have dropped the $m^4$ constant term.
The reduced Hamiltonian for the radial degree of freedom
 can be written  as
\begin{equation}
H=-\frac{1}{2}\frac{\partial^2}{\partial r^2}
+\frac{(N+2l-1)(N+2l-3)}{8r^2}
+4 \rho r^2
+\frac{4N\rho^{2}}{g}-\frac{Nm^{2}\rho}{g}
\end{equation}
where $l$ is a non-negative integer \cite{Chatterjee:1990se}.
Rescaling the radial coordinate $r\rightarrow N^{1/2}r$ leads to
a potential proportional to $N$
which is a function of both $r$ and $\rho$ and a kinetic term which is of order
$1/N$. It is easy to minimize the potential as a function of $r$; the final result,
after setting $l=0$,
is identical to the expression for $V_{eff}/N$  as a function of $\rho$
in the large-$N$ limit. % Revised
Thus we see that this radial formalism yields results for the ground state
energy equivalent to other approaches at leading order in the large-$N$ expansion. 
Alternatively, one can take the angular momentum quantum number $l$ to be of
order $N$. Elimination of the $r$ variable then leads to an effective potential of the form
\begin{equation}
V_{eff}=(N+2l)\sqrt{2\rho}
+\frac{4N\rho^{2}}{g}-\frac{Nm^{2}\rho}{g}
\end{equation}
which displays the $l$-dependent anomaly term first observed in \cite{Andrianov:2007vt}.
 It is also possible to show the equivalence of the radial formalism directly, without introducing the composite field $\rho$.
Note that
the radial approach demonstrates that the angular momentum term makes a  positive contribution to the ground state energy in the $PT$-symmetric case, 
exactly as it does in the Hermitian case.

\section{$PT$-symmetric field theory}

If we boldly apply the constraint field approach to a $PT$-symmetric
field theory with a $-g\left(\vec{\phi}^{2}\right)^{2}$ interaction
in $d$ dimensions, we obtain the effective potential\begin{equation}
V_{eff}/N=\frac{4\rho^{2}}{g}-\frac{m^{2}\rho}{g}+\frac{m^{4}}{16g}+\frac{1}{2}\int\frac{d^{d}k}{\left(2\pi\right)^{d}}\ln\left[k^{2}+8\rho\right].\end{equation}
Models of this type were rejected decades ago \cite{Coleman:1974jh}
because of stability
concerns at both the classical and quantum levels,
although there were early indications that
such theories were in fact sensible \cite{Andrianov:1981wu}.
Within the framework
of $PT$-symmetric models, such stability issues cannot be addressed
without a detailed understanding of the contours used
in functional integration.
However, it is straightforward to check that renormalization of $g$
in $d=4$ gives an asymptotically free theory, with beta function
$\beta=-g^{2}/2\pi^{2}$ in the large-$N$ limit. If $PT$-symmetric
scalar field theories exist in four dimensions and 
are indeed asymptotically free, the possible
implications for particle physics are large, and provide ample justification
for further work.

\vspace{0.5cm}
\footnotesize
\noindent
The authors would like to thank Carl M. Bender for many useful discussions;
we also thank Stefan Boettcher for providing the numerical 
data used in constructing Table 1. MCO gratefully acknowledges the support of the U.S.\ Department of Energy.

\end{document}